\documentclass[iop]{emulateapj}
\usepackage{graphicx}                                                         

\def\apj{Astrophys. J. }
\def\apjs{Astrophys. J., Suppl. Ser. }

\def\aap{A\&A}

\def\ga{\,\hbox{\hbox{$ > $}\kern -0.8em \lower 1.0ex\hbox{$\sim$}}\,}
\def\la{\,\hbox{\hbox{$ < $}\kern -0.8em \lower 1.0ex\hbox{$\sim$}}\,}

\def\kb{k_B}
\def\gcc{\,{\rm g}\,{\rm cm}^{-3}}

\def\kb{k_B}

\begin{document}

\title{A new equation of state for dense hydrogen-helium mixtures\\ II: taking into account hydrogen-helium interactions}

\author{Gilles Chabrier} 
\affil{
CRAL, Ecole normale sup\'erieure de Lyon, UMR CNRS 5574, 69364 Lyon Cedex 07,  France,\\
School of Physics, University of Exeter, Exeter, EX4 4QL, UK
}
\and 
\author{Florian Debras}
\affil{IRAP, Universit\'e de Toulouse, UMR CNRS 5277, UPS, Toulouse, France\\
}

\date{}

\begin{abstract}
In a recent paper (Chabrier et al. 2019), we have derived a new equation of state (EOS) for dense hydrogen/helium mixtures which covers the temperature-density domain  from solar-type stars to brown dwarfs and gaseous planets. This EOS is based on the so-called additive volume law and thus does not take
into account the interactions between the hydrogen and helium species. In the present paper, we go beyond these calculations by taking into account H/He interactions, derived from quantum molecular dynamics simulations. These interactions, which eventually lead to H/He phase separation,
become important at low temperature and high density, in the domain of brown dwarfs and giant planets. The tables of this new EOS are made publicly available.

%

\end{abstract}



\section{Introduction} 
\label{intro}

In a recent paper, Chabrier et al. (2019, Paper I) have derived a new equation of state (EOS) for dense hydrogen/helium mixtures which covers the temperature-density domain  from solar-type stars to gaseous planets. These calculations combine semi-analytic EOS models in the low density, low temperature molecular/atomic domain and   in the high-density, high-temperature fully ionized domain, respectively, and ab initio quantum molecular dynamics (QMD) calculations in the intermediate pressure dissociation and ionization domain. 
 This EOS adequately reproduces all existing experimental results, namely Hugoniot and isentropic shock experiments for pure H and He.
 It also agrees very well with  first principle numerical simulations for both the single elements and the mixture in most of the covered domain. Departure from
 the simulations, however, starts occuring below a temperature $T\lesssim 10^5$ K in a density domain $0.1\lesssim \rho \lesssim 10\gcc$, i.e. in the $\sim$Mbar regime (see e.g. Fig. 27 of Paper I). From the astrophysical point of view, this concerns the domain of so-called substellar objects: brown dwarfs and giant planets.
 
 This departure reflects the growing importance of the interactions between hydrogen and helium species, which eventually leads to a phase separation between these components. It is thus important to include these interactions in the H/He EOS in order to have a correct treatment of substellar object thermodynamic properties, thus  structure and evolution. The EOS derived in Paper I is based on the so-called additive volume law (AVL) approximation and thus does not take into account H/He interactions (see eqns (8)-(11) of Paper I). It is the aim of this paper to go beyond this limitation by including H/He interactions, derived from
 ab initio QMD calculations (Militzer \& Hubbard 2013, MH13). 

The paper is organized as follows. In \S2, we derive an "effective" EOS for pure hydrogen based partly on the one derived for H/He by MH13 in the relevant temperature-pressure domain. In \S3, we use this effective hydrogen EOS to derive a revised H/He EOS for various helium mass fractions and we carry out extensive comparisons of this new H/He EOS with the MH13 QMD calculations. The tables are presented in this same section while section 4 is devoted to the conclusion.

\section{An effective Hydrogen equation of state}
\label{EOSH}

\begin{figure}
\includegraphics[width=12cm,width=7cm,angle =-90]{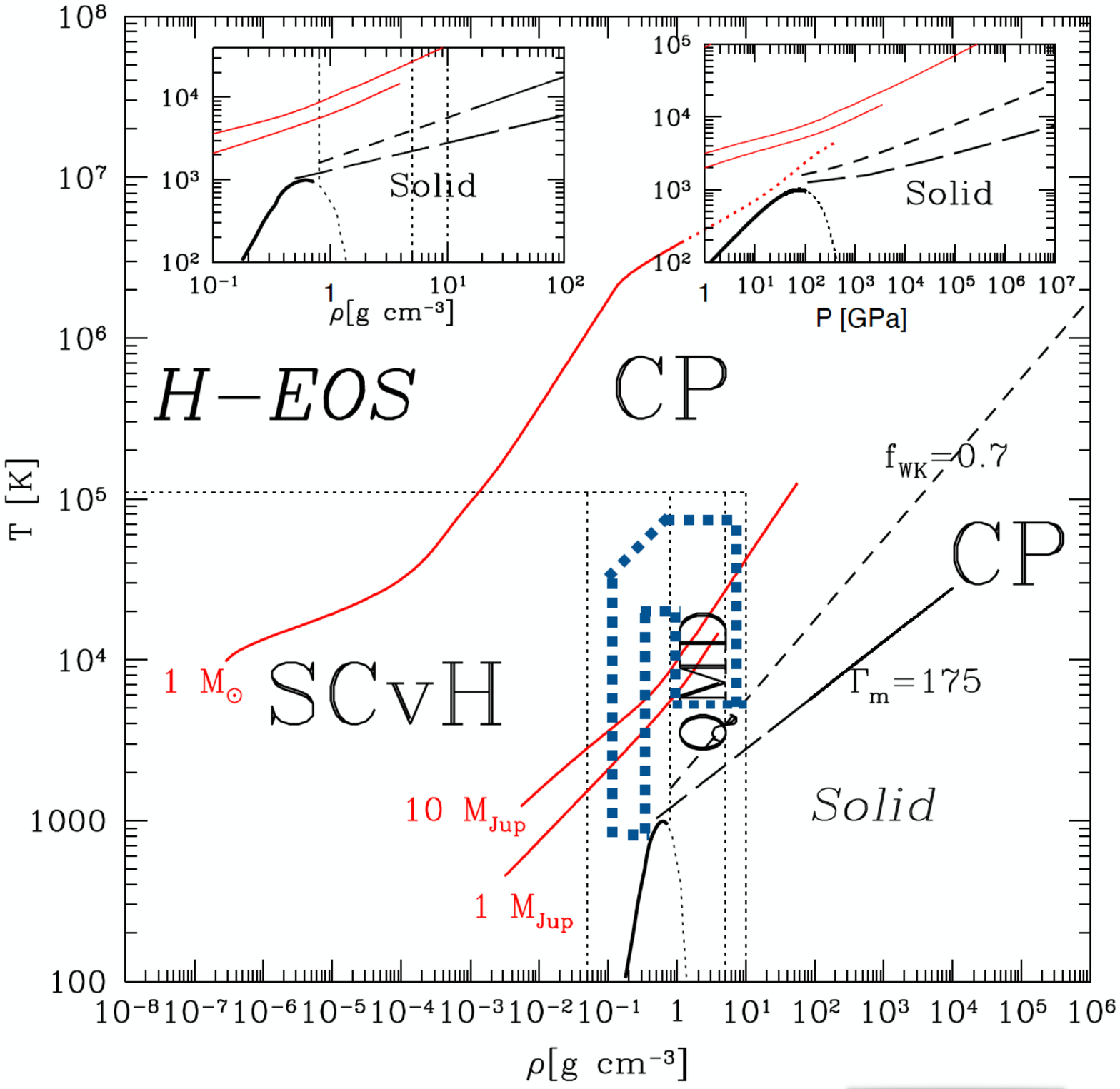}
\caption{Same figure as Fig. 1 of Chabrier et al. (2019, Paper I) to illustrate the temperature-density domain of the H EOS. The (blue) dashed lines in the center of the figure delimitate the domain covered by the Militzer \& Hubbard (2013, MH13) EOS simulations.
All other symbols are the same as in Fig. 1 of Paper I. }
\label{fig1}
\end{figure}

To derive the effective H EOS based on the QMD calculations by MH13 in the pressure ionization regime, we proceed as in Debras \& Chabrier (2019). The procedure is similar to the one used in Miguel et al. (2016) except that, in contrast to these latter,
 it does take into account the mixing entropy contribution in the H/He MH13 EOS (eqns.(\ref{newS}) below). As discussed below, such a contribution is not negligible and affects the resulting EOS. 
 The H/He EOS table derived by MH13 is based on QMD simulations and provides the internal energy and pressure for 391 temperature-density points and a typical mass fraction $Y=0.245$ within the domain $T\in [10^3,8\times 10^4]$ K, $\rho \in [\sim 0.2, 9.0]\gcc$. For 131 of these points, 
 an ab initio thermodynamic integration technique was performed to provide the Helmholtz 
 free energy and the entropy (see MH13 for details).
 Figure \ref{fig1} shows the corresponding domain covered by the MH13 EOS in the pure hydrogen diagram.
 
 We  first calculate at each $P$-$T$ point calculated by MH13 the corresponding hydrogen density:

\begin{eqnarray}
\frac{1}{\rho_{\mathrm{MH13}}} &=& \frac{X_\mathrm{MH13}}{\rho_H}+ \frac{Y_\mathrm{MH13}} {\rho_{\mathrm{He}}}\\
 \Rightarrow  \frac{1}{\rho_{\mathrm{H}}}&=&\frac{1}{X_\mathrm{MH13}}\left(\frac{1}{\rho_{\mathrm{MH13}}} - 
\frac{Y_\mathrm{MH13}} {\rho_{\mathrm{He}}} \right), 
\label{newrho}
\end{eqnarray}
where $\rho_H$ is the (sought) mass density for pure hydrogen, $X_\mathrm{MH13}=0.754$, $Y_\mathrm{MH13}=0.246$ 
the mass fractions of hydrogen and helium in the MH13 simulations ($N_\mathrm{H}=220$ and $N_\mathrm{He}=18$ the numbers of H and He particles, respectively), $\rho_{\mathrm{MH13}}$ the density derived from MH13 by spline procedures;
$\rho_{\mathrm{He}}$ the helium density in the Chabrier et al. (2019) EOS. This reduced table, which covers only the limited $T$-$P$ domain explored by MH13, is
then combined by spline procedures with the pure H EOS of Paper I in the remaining $T$-$P$ domain covered by this latter, where the AVL becomes valid.

\bigskip

Similarly, the pure hydrogen entropy is obtained from 
\begin{eqnarray}
S_H =  \frac{1}{X_\mathrm{MH13}}\left(S_\mathrm{MH13} - Y_\mathrm{MH13}S_{\mathrm{He}} - S_\mathrm{mix}\right),
\label{newS}
\end{eqnarray}
with  $S_\mathrm{MH13}$ the splined specific entropy from MH13,
$S_{\mathrm{He}}$ the helium specific entropy in the Chabrier et al. (2019)  EOS, all at the same ($P$,$T$), and $S_\mathrm{mix}$ is 
the mixing specific entropy. 

As mentioned above and will be shown in \S3, $S_\mathrm{mix}$ can be of the order of several percents and thus should not be neglected to derive the effective pure H EOS.
Strictly speaking, the derivation of the effective pure H EOS is not fully consistent. Indeed, even though the MH13 calculations take into account the non-ideal H/He mixing entropy, and then so does $S_\mathrm{mix}$ in eqn. (\ref{newS}), 
eqn. (\ref{newrho}) is based on the AVL mixing equations and thus does not properly take into account the excess volume/density of {\it both species} at constant pressure, $P\Delta (V_\mathrm{H}+V_\mathrm{He})$. In our procedure, all the excess volume is somehow included in the pure hydrogen one. 
As will be examined in detail in \S\ref{calc}, in the region of maximum departure from the AVL, T$\sim 5,000$-10,000 K, P$\sim 10$-100 GPa, i.e. $\rho \sim 0.1$-1$\gcc$, this excess volume remains modest, $\lesssim 4\%$, consistent with the value found for the H/He mixture in detailed calculations (see e.g. Fig. 17 of Vorberger et al. 2007). An other obvious limitation of the present calculations is that the effective hydrogen entropy corresponds  to one single value of $Y=Y_\mathrm{MH13}=0.246$. 
The derivation of fully consistent H/He EOSs, however, would require
carrying out ab-initio QMD simulations for various values of $Y$ over
a large enough ($T,\rho$) domain to smoothly reach the domains where the AVL becomes valid. This represents quite a heavy numerical task. 
Given the lack of such calculations, we will stick to the present derivation, based on a combination of the present effective hydrogen EOS and our helium EOS, to derive EOS tables  for any helium mass fraction, as required for giant planet interior structure calculations (see e.g. Debras \& Chabrier 2019). 
We will come back to this point in \S3.1.

As mentioned in Paper I, the EOS is calculated initially in a $T$-$\rho$ domain, the thermodynamic variables used in QMD or PIMC calculations (carried out in the canonical ensemble),
 and then transformed into a $T$-$P$ one by bicubic interpolation procedures. 
 It is inevitable that these successive  procedures lead to numerical errors, notably for second derivative variables. We will examine the impact of these errors in the following section by comparing our results with the original ab initio calculations.
 A first estimate of these errors is displayed in Fig. 1 of Debras \& Chabrier (2019). These authors used at that time a subset of the present EOS to derive Jupiter interior models that fullfil the JUNO (Iess et al. 2018) and GALILEO (Wong et al. 2004) constraints. This figure shows the relative error on the density between the present EOS and the MH13 one, as well as with Miguel et al. (2016). 
As seen in this figure, for Jupiter interior conditions, the relative error on the density, $(\rho - \rho_{\mathrm{MH13}}) / \rho_{\mathrm{MH13}}$, between the present and MH13 calculations for Jupiter interior conditions is always $<1.0 \%$, which is less than the numerical error in the MH13 simulations.
In contrast the error for the Miguel et al. calculations reaches several percents on the density. The error on the entropy is shown to be on the same amount for
the present EOS ($<0.5\%$), whereas it reaches up to $30\%$ for Miguel et al. (2016).

We stress that this revised pure hydrogen EOS  is only intended to be used for the H/He mixtures to somehow take into account the H-He interactions. It must {\it not} be used as a pure H EOS. For this latter case, one must use the H EOS derived in Paper I, based in part on various ab-initio simulations for pure H. For illustrative purposes, Fig. 2 compares the  Hugoniot curves for deuterium calculated with the present H EOS and with the one calculated in
Paper I (Chabrier et al. 2019) with the experimental data of Knudson \& Desjarlais (2017). Even though the differences remain small over most of the domain
probed by the experiments, we see that the agreement with the data in the region of maximum compression is better with the H EOS calculated in Paper I.
We carried out similar comparisons for the hydrogen and deuterium Hugoniot experiments of Brygoo et al. (2015), for their various precompressed initial conditions (see Paper I).
We found out that under the conditions probed by these experiments, the difference between the present and 2019 EOS is even more modest. This gives  confidence in the way we have calculated the present
revised H EOS.

\begin{figure}
\includegraphics[width=\linewidth]{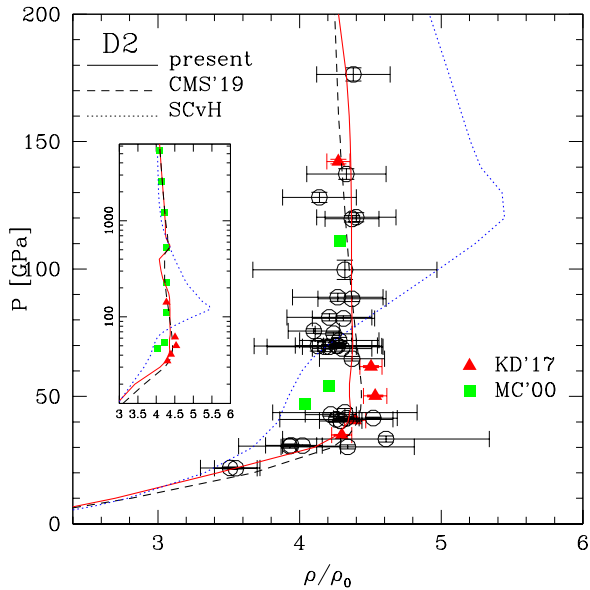}
\vspace{-2cm}
\caption{Shock pressure vs density along the deuterium Hugoniot curve. Solid triangles: results by Knudson \& Desjarlais (2017) for initial temperature
and density $T_0=20$ K and $\rho_0=0.167\gcc$, respectively. Empty circles: reanalyzed shock data obtained from various experiments rescaled to the
same initial density (data from Knudson \& Desjarlais (2017)). Solid squares: PIMC calculations of Militzer \& Ceperley (2000). Solid (red) line: present calculations; dashed (black) line: EOS of Paper I (Chabrier et al. 2019); dotted (blue) line: SCvH EOS (Saumon et al. 1995).}
\label{fig2}
\end{figure}

\section{The Hydrogen/Helium mixture equation of state}
\label{EOSHHe}

\subsection{Calculation of the H/He EOS}
\label{calc}

As mentioned in the conclusion of Paper I, while thermodynamic quantities for the H/He mixture calculated with the "additive volume law" (AVL) compare quite well with ab-initio calculations for isentropes larger than about 12 $\kb/e^-$ ($\sim 13\,\kb/at$ ), i.e. about $9\times 10^{-2}$ MJ kg$^{-1}$ K$^{-1}$, the disagreement becomes noticeable below this value. This stems from the increasing interactions between hydrogen and helium atoms for $T\lesssim {\rm a\,\, few}\,\,10^4$ K and $\rho \gtrsim 0.1\gcc$ (see Fig. 27 of Paper I). Such a domain encompasses essentially all substellar objects, brown dwarfs and giant planets, and is thus of high astrophysical interest.

The revised H/He EOS and the corresponding thermodynamic quantities are calculated exactly as in \S4 of Paper I. Figures 3-9 display exactly the same comparisons
with the MH13 simulations as in Figures 19-27 of Paper I. 

\begin{figure}
\includegraphics[width=\linewidth]{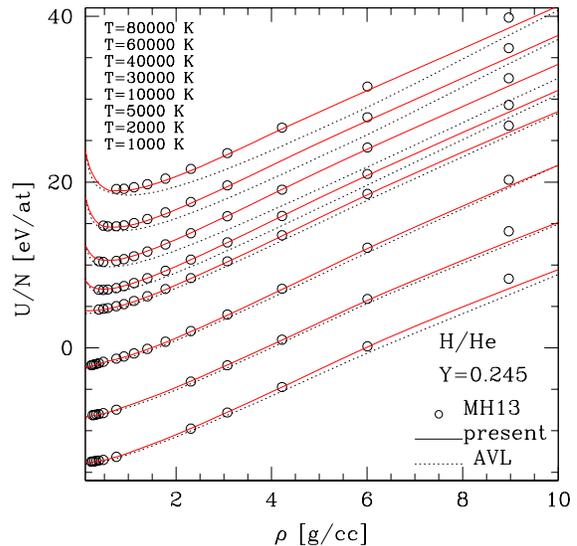}
\vspace{-2cm}
\caption{Internal energy per atom vs density for several isotherm calculations by Militzer \& Hubbard (2013, MH13) (as labeled in the figure) for $Y=0.245$, compared with the present EOS and the one based on the AVL (Paper I), respectively.
For all curves the zero of energy is the same as in MH13.
For sake of clarity, however, curves have been arbitrarily moved upward or downward by constant shifts.} 
\label{fig3}
\end{figure}

\begin{figure}
\includegraphics[width=\linewidth]{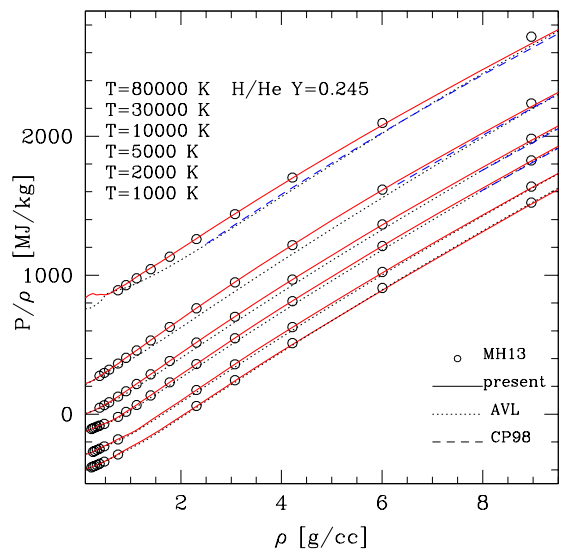}
\vspace{-2cm}
\caption{Same as Fig. \ref{fig3} for the non-ideal pressure $P/\rho$. The (blue) dashed lines correspond to the fully ionized model of Chabrier \& Potekhin (1998).} 
\label{fig4}
\end{figure}

Figures \ref{fig3} and \ref{fig4} portray the internal energy par atom and the non-ideal pressure, $P/\rho$, for several isotherms, for $Y=0.245$,
the helium mass fraction used in MH13. We note the excellent agreement between the present H/He EOS and the MH13 simulations, whereas
the AVL-based EOS (Paper I) departs noticeably from these latter in the $T$-$\rho$ domain displayed in the figures. As seen in Fig. \ref{fig4},
for $\rho \ga 8\gcc$, we recover the fully ionized plasma model of Chabrier \& Potekhin (1998).

\begin{figure}
\includegraphics[width=\linewidth]{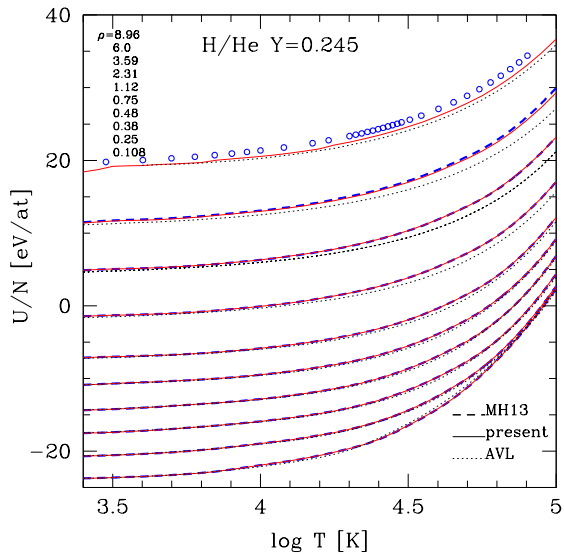}
\vspace{-2cm}
\caption{Internal energy per atom vs temperature for several isochore calculations by Militzer \& Hubbard (2013, MH13) (as labeled in the figure), compared with the present and AVL results.
For all densities, the MH13 values are the ones given by their fit except for $\rho=8.96\gcc$, which is out the range of validity
of the fit and for which the empty circles are their simulation data points.
Solid lines: present calculations; blue long-dashed lines and empty circles: MH13; black dotted lines: AVL (Paper I).
For all curves the zero of energy is the same as in MH13.
For the sake of clarity, however, curves have been moved arbitrarily upward or downward by constant shifts.} 
\label{fig5}
\end{figure}

\begin{figure}
\includegraphics[width=\linewidth]{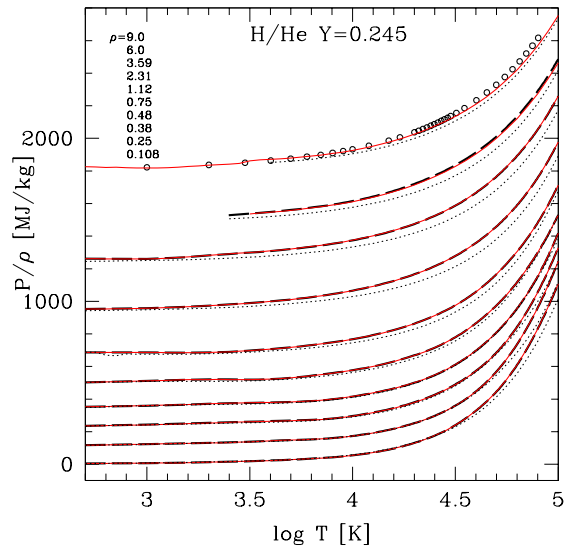}
\vspace{-2cm}
\caption{Same as Fig. \ref{fig5} for the non-ideal pressure $P/\rho$. 
 For the sake of clarity, curves have been shifted upward arbitrarily by constant shifts.}
\label{fig6}
\end{figure}

Figures \ref{fig5} and \ref{fig6} portray similar comparisons for various isochores. Again we note the excellent agreement between MH13 and the present calculations and the departure of the AVL EOS for $T\ga 10^4$ K, $\rho \ga 1\gcc$. 

\begin{figure}
\includegraphics[width=\linewidth]{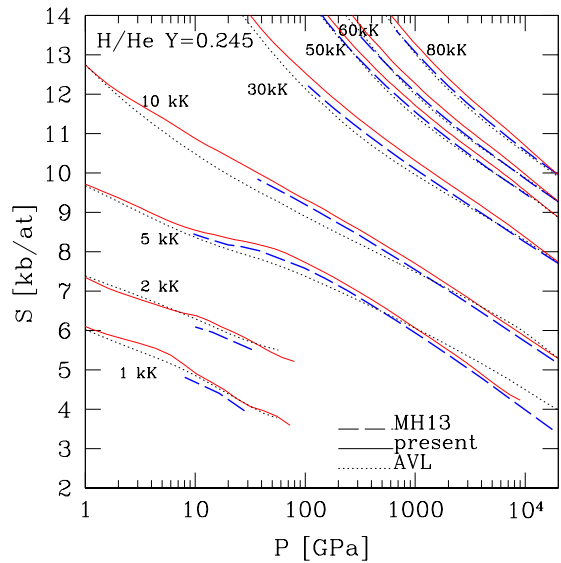}
\vspace{-2cm}
\caption{Entropy vs density for several isotherm calculations by Militzer \& Hubbard (2013) (as labeled in the figure), compared with the present and AVL results, respectively. 
Same labeling as in the previous figures.} 
\label{fig7}
\end{figure}

\begin{figure}
\includegraphics[width=\linewidth]{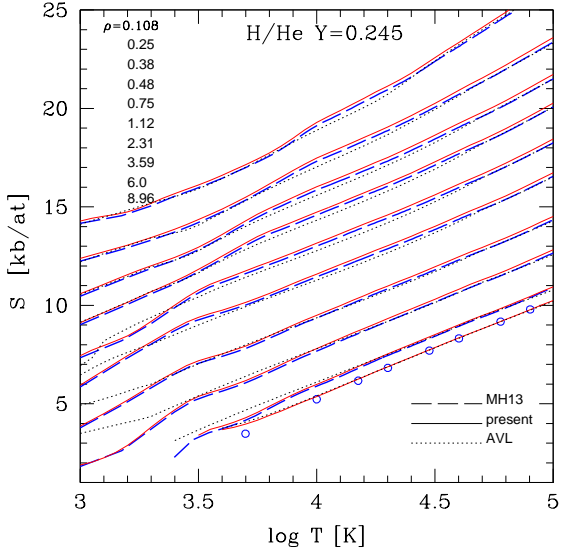}
\vspace{-2cm}
\caption{Entropy vs temperature for several isochore calculations by Militzer \& Hubbard (2013) (as labeled in the figure), compared with the present and AVL results, respectively. Same labeling as in the previous figures. Careful: for sake of clarity, each curve from $\rho=3.59\gcc$ to $\rho=0.108\gcc$ has been shifted upward by 1 $k_b/$atom w.r.t. to the
immediately higher-density one.} 
\label{fig8}
\end{figure}

Figures \ref{fig7}  and \ref{fig8} show the same comparisons for the entropy. 
Of noticeable interest is the excellent agreement between the present calculations and the MH13 simulations for the $T=5000$ K and $T=10,000$ K isotherms,
for which departure from the AVL approximation is the largest over essentially the entire density range displayed in the figure, highlighting
the importance of H-He interactions in this domain. This brings confidence in our procedure, even though approximated, to take into account the {\it non-ideal volume and
entropy of mixing}, naturally included in the MH13 simulations. The difference between the present or MH13 calculations and the ones based on the AVL amounts to $\lesssim 5\%$ 
of the total entropy in the 5,000-10,000 K, 10-100 GPa (i.e. $\sim 0.1$-1$\gcc$) $T$-$P$ maximum departure range, of which more than 3\% stems from the ideal entropy contribution (eqn.(11) of Paper I).  
We do not show comparisons for the free energy (Figs. 23 and 24 of Paper I) as the present and AVL calculations are essentially indistinguishable for this quantity,
an obvious consequence of the compensating increased values of $U$ and $S$ on $F=U-TS$ when taking into account H/He interactions, an effect already found
for the fully ionized mixtures (Chabrier \& Ashcroft 1990).

\begin{figure}
\includegraphics[width=\linewidth]{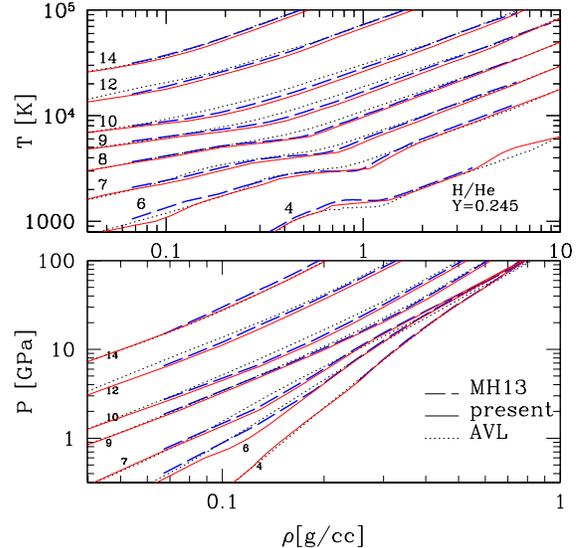}
\vspace{-2cm}
\caption{Temperature and pressure profiles for a series of adiabats as labeled in the figure in $\kb/e^-$ ($=(\kb/{\rm atom})/1.076$ for the present $Y$ value) for the MH13 mass fraction
of helium ($Y=0.245$). Solid lines: present calculations; blue long-dashed lines: MH13; black dotted lines: AVL.} 
\label{fig9}
\end{figure}

We now carry out similar comparisons along isentropic profiles. This is illustrated in Figure \ref{fig9}.
As mentioned in Paper I and seen in the figure, non-ideal effects between the H and He species become noticeable for entropies
$S\lesssim 12\kb/e^-$ ($\lesssim 13\,\kb/at$ ), i.e.  $\lesssim 9\times 10^{-2}$ MJ kg$^{-1}$ K$^{-1}$. Here again, we note the  good agreement between the present calculations and the MH13 simulations. The new calculations yield cooler and denser profiles for a given pressure than the calculations based on the AVL approximation, yielding more compact adiabatic structures.
Of noticeable interest is the excellent agreement for the 7$\kb/e^-$, characteristic of Jupiter
and Saturn interiors. This assesses the validity of the EOS used in the calculations of Debras \& Chabrier (2019). Note that according to the calculations of Militzer \& Hubbard
(2013), adiabats for $S\lesssim 6\kb/e^-$ pass through the H/He immiscibility region where the mixed state described by the simulations is no longer thermodynamically stable.

All these comparisons bring confidence in the present calculations of an EOS for H/He mixture beyond the AVL approximation, i.e. which includes,
even though in an approximated way, the impact of H-He interactions on the thermodynamic properties of the mixture. Of course, one must remain cautious about the accuracy of the present EOS for other helium mass fractions. Only a comparison with ab-initio simulations for various $Y$ values could verify the agreement. However, the present test at least assesses the validity of our H/He
EOS for substellar objects of solar or near solar compositions.

\subsection{Form of the EOS tables}

All tables of the new EOS have the same rectangular form as for the ones presented in Paper I, and are given either
with ($T,\rho$) or ($T,P$) independent variables. Because of the various spline procedures either in the interpolated regimes between different calculations
or in the ($T,P$) to ($T,\rho$) transformations, the EOS, unfortunately, suffers from the same unphysical numerical oscillations in some domains, notably for the second derivatives.
Note that the present tables have different limits for the density than the ones of Paper I, namely:

\begin{eqnarray}
2.0 &\le& \log\, T\le 8.0,\nonumber \\
 -9.0&\le& \log\, P\le +13.0, \nonumber \\
 -6.0 &\le& \log\, \rho \le +6.0,
\end{eqnarray}
with grid spacings $\Delta \, \log T=0.05$, $\Delta \, \log P=0.05$, $\Delta \, \log \rho=0.05$, i.e.  121 isotherms, each with 441 values of $P$ or 241 values of $\rho$, and $T$ in K, $P$ in GPa, $\rho$ in $\gcc$. Indeed, calculations for $ \log\, \rho \le -6.0\gcc$ were found to yield spurious results. In any case, the concept of equilibrium
thermodynamic quantities for such low densities becomes of dubious validity. The domains of validity of the EOS remain the same as the ones defined in Paper I,
due to the onset of ion quantum effects or crystallization (see Figs. 1 and 16 of Paper I and discussions therein).

The tables have been calculated for 3 helium mass fractions, namely $Y_\odot=0.275$, the helium cosmogonic abundance, $Y_{eq}=0.275+0.017=0.292$,
the value used in the Lyon group solar composition brown dwarf evolutionary calculations (Baraffe et al. 2003, Chabrier et al. 2000, Phillips et al. 2020), and
$Y_{eq}=0.28+0.017=0.297$, the value used in evolutionary calculations 
of solar composition low-mass stars (Baraffe et al. 2015)\footnote{$Y=0.28$ is the helium fraction required to recover the present radius
and luminosity of the Sun in the Baraffe et al. (2015) models. It must be kept in mind that the exact abundance of the Sun is not known.}.
 The 'equivalent' helium fraction, $Y_{eq}=Y_\odot+Z_\odot$, with $Z_\odot=0.017$ (Asplund et al. 2009), is the simplest (even though crude)
way to take into account the (small) contribution of heavy elements to the EOS.\

{\underline {\it A word of caution}}: In the {\it stellar} range, non ideal H/He contributions are basically inconsequential.
Therefore, in this domain, users should use the AVL EOS tables (Chabrier et al. 2019) 
  in order to avoid further numerical issues due to spline procedures.
The present tables should be used essentially in the {\it brown dwarf and giant planet}  domain, where 
H/He interactions start to play a role (see Chabrier et al. 2021, in preparation).

\section{Conclusion}

In this paper, we have derived a new equation of state for hydrogen and helium mixtures which incorporate the results of the simulations performed by Militzer \& Hubbard (2013) for $Y=0.245$. In order to do so we first calculate an effective EOS for pure hydrogen and combine this latter with our pure helium EOS (Paper I) to derive the one
for the mixture. This allows us to somehow take into account the impact of the H-He interactions upon the thermodynamic quantities of the mixture. Although the procedure
is not entirely satisfactory, due to the lack of a proper determination of the non-ideal volume and entropy of mixing from the simulations, comparisons between our calculations
and the ones by MH13 for different thermodynamic quantities for various isotherms, isochores and isentropes in the region of pressure dissociation/ionization show an
excellent agreement. Although being not a genuine proof of validity of our procedure for any mixture composition, this at least assesses its validity for solar-like H/He compositions.
The comparison between MH13 simulations and the present calculations for cool isentropes, in particular, for which H/He interactions become dominant, yielding eventually a
demixing process, show an excellent agreement. The impact of the H/He interactions on these isentropic (T,P) structures is found to be substantial, yielding cooler and denser
 profiles, as already found in MH13. This assesses the validity of the present H/He EOS in the domain of substellar objects, brown dwarfs and giant planets, to
derive more correct structure profiles and cooling sequences. This will be examined in more details in a forthcoming paper (Chabrier et al., 2021, in preparation). 

An obvious weakness of the present EOS calculations is that they rely entirely on one single set of ab-initio simulations, carried out for one single helium composition.
A point of concern is that various ab initio (quantum molecular dynamics or path integral) calculations still show differences (see discussion in Paper I). As pointed out in the conclusion of Paper I (see also Ramakrishna et al. 2020), 
the different exchange-correlation functionals used in the calculations for liquid hydrogen, for instance, yield pressures that can differ by as much as $\ga 10\%$ in the domain of pressure dissociation and ionisation.
Whether these differences will have a strong or small impact on the present EOS calculation and its astrophysical applications remains to be determined. More ab-initio simulations, possibly with different numerical tools, over a larger range of helium mass fractions,
are strongly needed to clarify this issue and definitely validate or not the present EOS calculations for different H/He mixture compositions. This is crucial  notably to determine the internal structure and composition of 
our Solar System planets. 

Calculations for other values of $Y$ can be calculated as described by eqns.(8)-(11) of Paper I, using the present revised H EOS and the He EOS of Paper I (requests should be sent to G. Chabrier).
However, we stress again that the present revised H EOS should be used only for this purpose and must not be used as a pure hydrogen EOS, notably when comparing with ab initio simulations or high pressure experiments for pure hydrogen or deuterium.

The present revised H/He EOS tables for $Y=0.275$, $Y=0.292$, and $Y=0.297$, as well as the ones for pure H and pure He (Chabrier et al., 2019) are available in a .tar.gz package with this article. The present data, along with the 2019 dataset, can also be obtained at http://perso.ens-lyon.fr/gilles.chabrier/DirEOS

\acknowledgments
This work has been partly supported by the Programme National de Plan\'etologie (PNP).
\\

\centerline{REFERENCES}
\noindent 
Asplund, M., et al., 2009, \araa, 47, 481\\
Baraffe, I., Chabrier, G., Barman, T., Allard, F., \& Hauschild, P., 2003, \aap, 402, 701\\
Baraffe, I., Homeier, D., Allard, F., \& Chabrier, G., 2015, \aap, 577, 42\\
Brygoo, S.  et al., 2015, J. Appl. Phys., 118, 195901 \\
Chabrier, G., \& Ashcroft, N., 1990, \pra, 42, 2284   \\
Chabrier, G., \& Potekhin, A., 1998, \pra, 58, 4941   \\
Chabrier, G., Baraffe, I., Allard, F., \& Hauschild, P., 2000, \apj, 542, 464\\
Chabrier, G., Mazevet, S. \& Soubiran, F., 2019, \apj, 872, 51 \\
Debras, F. \& Chabrier, G., 2019, \apj, 872, 100 \\
Iess, L. et al., 2018, Nature, 555, 220 \\
Knudson, M., \& Desjarlais, M., 2017, \prl, 118, 035501  \\
Miguel, Y., Guillot, T. \& Fayon, L., 2016, \aap, 596, 114 \\
Militzer, B., \& Hubbard, W., 2013, \apj, 774, 148   \\
Phillips, M., Trembin, P., Baraffe, I. et al., 2020, \aap, 637, 38 \\
Ramakrishna,  K.,  Dornheim, T. \&  Vorberger, J., 2020, \prb, 101, 195129 \\
Saumon, D., Chabrier,G., \& van Horn, H., 1995, \apjs, 99, 713  \\
Vorberger, J. et al., 2007, \prb, 75, 4206 \\ 
Wong, M.H. et al., 2004, Icarus, 171, 153

\end{document}